\documentclass{aa}
\usepackage{txfonts}
\usepackage{graphicx}
\usepackage{subcaption}
\usepackage[colorlinks=true,allcolors=blue]{hyperref}
\usepackage{gensymb}

\DeclareUnicodeCharacter{5415}{}  
\DeclareUnicodeCharacter{5EFA}{}  
\DeclareUnicodeCharacter{4F1F}{}  

\begin{document} 

\authorrunning{Mountrichas et al.}
\titlerunning{AGN in Dwarf Galaxies}

\title{Fueling, Evolution, and Diversity of AGN in Dwarf Galaxies: Insights from Star Formation and Black Hole Scaling Relations}

\author{G. Mountrichas\inst{1}, M. Siudek\inst{2,3}, F. J. Carrera\inst{1}}
          
     \institute {Instituto de Fisica de Cantabria (CSIC-Universidad de Cantabria), Avenida de los Castros, 39005 Santander, Spain.
              \email{gmountrichas@gmail.com}
              \and
       Instituto de Astrof\'{\i}sica de Canarias, V\'{\i}a L\'actea, 38205 La Laguna, Tenerife, Spain\label{aff1}
\and
Instituto de Astrof\'isica de Canarias (IAC); Departamento de Astrof\'isica, Universidad de La Laguna (ULL), 38200, La Laguna, Tenerife, Spain\label{aff2}}

\abstract{We investigate the star formation activity and black hole scaling relations in a sample of 1\,451 active galactic nuclei (AGN) hosted by dwarf galaxies ($\log(M_\star/M_\odot) < 10$) at $\rm 0.5 < z < 0.9$, drawn from the VIPERS survey. The sample comprises Seyferts and LINERs identified through emission-line diagnostics, as well as IR-selected AGN based on WISE mid-infrared colors. Using the normalized star formation rate (SFR$_{\mathrm{norm}}$), defined as the ratio of the SFR of a galaxy hosting an AGN to the median SFR of star-forming galaxies of similar stellar mass and redshift, we compare AGN hosts to a control sample of non-AGN star-forming galaxies. We examine how SFR$_{\mathrm{norm}}$ varies with AGN power (L[OIII]), black hole mass (M$_{\mathrm{BH}}$), local environment and stellar population age. We also analyze the M$_{\mathrm{BH}}$–M$_\star$ relation and the evolution of the M$_{\mathrm{BH}}$/M$_\star$ ratio, incorporating comparisons to X-ray AGN and high-redshift quasars ($z > 4$). We note that black hole masses have been estimated from narrow-line diagnostics, which introduce significant scatter and may carry substantial uncertainties for individual sources, though they remain useful for identifying statistical trends. Our key findings are:
(i) All AGN populations show suppressed star formation at low AGN luminosities, with SFR$_{\mathrm{norm}}$ rising above unity at different luminosity thresholds depending on AGN type.  
(ii) LINERs show flat SFR$_{\mathrm{norm}}$ trends with M$_{\mathrm{BH}}$, remaining broadly consistent with unity. Seyferts display a mild increase with M$_{\mathrm{BH}}$, while IR AGN show a more pronounced positive trend.  
(iii) LINERs exhibit older stellar populations than Seyferts.  
(iv) At fixed stellar mass, Seyferts host more massive black holes than LINERs, with IR AGN falling in between.  
(v) The M$_{\mathrm{BH}}$/M$_\star$ ratio is elevated relative to local scaling relations and remains approximately constant with redshift over $0.5 < z < 0.9$, in agreement with high-$z$ AGN measurements.  
(vi) The ratio decreases with stellar mass up to $\log(M_\star/M_\odot) \sim 11$, beyond which it flattens toward values consistent with those of local, inactive galaxies; this trend is clearest for Seyferts and IR AGN, while LINERs show no clear dependence.
These results suggest that AGN in dwarf galaxies follow diverse evolutionary pathways, shaped by gas availability, feedback, and selection effects.

}

   
\maketitle  

\section{Introduction}

Supermassive black holes (SMBHs) are widely accepted as key components influencing the evolution of their host galaxies through energetic feedback processes associated with active galactic nuclei (AGN). AGN activity, fueled by accretion onto SMBHs, can inject substantial amounts of energy into their surroundings, affecting star formation, gas distribution, and the overall growth and quenching of galaxies \citep[e.g.,][]{Silk1998, Kormendy2013}.  Understanding the role and impact of SMBHs in galaxy evolution is thus fundamental for building comprehensive galaxy formation models.

While AGN activity has been extensively studied in massive galaxies, dwarf galaxies, typically defined as having stellar masses $\log(M_\star/M_\odot) \lesssim 10$ \citep[e.g.,][]{Reines2013, Mezcua2016, Greene2020} have recently attracted significant attention as valuable laboratories for understanding early phases of galaxy evolution and SMBH formation \citep[e.g.,][]{Penny2018, ManzanoKing2019, Birchall2020}. Dwarf galaxies are important because they represent the most abundant galaxy population in the Universe \citep[e.g.,][]{Baldry2012} and are believed to host intermediate-mass black holes (IMBHs; $10^3 - 10^5\,M_\odot$) that may provide crucial insights into the seed black holes that evolved into SMBHs in more massive galaxies \citep[e.g.,][]{Volonteri2010, Mezcua2017, Greene2020}.

In recent years, a growing number of observational studies have successfully identified AGN in dwarf galaxies \citep[e.g.,][]{Reines2013, Pardo2016, Baldassare2016, Baldassare2020, Mezcua2016, Mezcua2018, Mezcua2020, Kaviraj2019, Greene2020, Manzanoking2020, Polimera2022, Siudek2023a, Pucha2025}. These studies have revealed several key trends: AGN in dwarf galaxies are preferentially found in environments that support gas-rich interactions or mergers \citep[e.g.,][]{Kaviraj2019, Mezcua2020, Erostegui2025}, display enhanced central star formation rates compared to inactive dwarfs \citep[e.g.,][]{Reines2013, Polimera2022}, and may significantly affect the chemical and morphological evolution of their hosts \citep{Mezcua2018}. Theoretically, recent simulations suggest that AGN feedback in dwarf galaxies can play a pivotal role in regulating their evolution—affecting gas retention, quenching star formation, and reshaping their growth pathways \citep[e.g.,][]{Koudmani2024, ArjonaGalvez2024}. These models highlight the sensitivity of dwarf systems to even low-level SMBH activity, and predict that such feedback can leave long-lasting imprints on their star formation and structural properties.

A fundamental approach to investigate the SMBH–galaxy co-evolution is by comparing the star formation rates (SFRs) of galaxies hosting AGN to those without active nuclei. Such comparative analyses have been extensively conducted for massive galaxies, revealing complex trends of star formation enhancement or suppression correlated with AGN luminosity and host galaxy properties \citep[e.g.,][]{Rosario2012, Santini2012, Mullaney2015, Masoura2018, Bernhard2019, Florez2020, Masoura2021, Mountrichas2021b, Mountrichas2022a, Mountrichas2022b, Pouliasis2022, Koutoulidis2022, Torbaniuk2024, Mountrichas2024d, Mountrichas2024c, Mountrichas2024e, Cristello2024, Zhang2025, Kondapally2025}. In the context of dwarf galaxies, recent studies have found that AGN feedback does not significantly impact the star formation of their hosts \citep{Siudek2023a}. A detailed multiwavelength of the dwarf galaxy NGC 4395 suggest that some AGN-hosting dwarfs may exhibit centrally enhanced star formation, indicating that low-luminosity AGN activity can co-exist with, or even promote, star formation on sub-galactic scales \citep{Nandi2023}.

Another critical avenue for exploring the connection between SMBHs and their host galaxies is through the correlation between black hole mass (M${_\mathrm{BH}}$) and stellar mass (M$_\star$), extensively studied in massive galaxies up to redshift $z \sim 2$ \citep[e.g.,][]{Magorrian1998, Ferrarese2000, Jahnke2009, Kormendy2013, Reines2015, Sun2015, Suh2020, Setoguchi2021, Mountrichas2023b}. For dwarf galaxies, however, this correlation is less well established due to the intrinsic observational challenges and the scarcity of robust SMBH measurements. Recent studies addressing this issue indicate that dwarf galaxies tend to host relatively massive black holes for their stellar mass, implying that SMBH growth may precede or outpace stellar growth in low-mass systems \citep[e.g.,][]{Mezcua2023, Mezcua2024, Greene2024, Sun2025}. Notably, similar elevated black hole-to-stellar mass ratios have been observed at high redshift ($\rm z > 4$), hinting at potentially analogous growth mechanisms or evolutionary pathways operating in both early-universe galaxies and local dwarf populations \citep[e.g.,][]{Sun2025b, Ding2023, Yue2024}.

In this paper, we present a detailed analysis of the interplay between star formation activity, black hole mass, and host galaxy properties in a large, systematically selected sample of dwarf galaxies hosting AGN drawn from the VIPERS survey within the redshift range $0.5 < z < 0.9$. Utilizing optical and infrared diagnostics, we identify Seyferts, LINERs, and IR-selected AGN, alongside a carefully constructed control sample of star-forming galaxies matched in stellar mass, color, and redshift. We investigate how AGN activity correlates with star formation and local environment, and we extend our analysis to include comparisons with X-ray selected AGN populations and high-redshift sources to provide a comprehensive view of SMBH–galaxy co-evolution across a wide range of galaxy masses and cosmic epochs.

\section{Data}
\label{sec_data}

In this section, we briefly describe the parent dataset, the derived physical parameters, and the methods used to compute them. A comprehensive overview of the data and analysis procedures can be found in \citet{Siudek2023a} (see their Section 2). For completeness, we provide a concise summary below.

\subsection{The parent sample}
\label{sec_parent_sample}

The galaxy sample analyzed in this study is drawn from the VIMOS Public Extragalactic Redshift Survey \citep[VIPERS;][]{Scodeggio2018}, a large spectroscopic campaign conducted with the VIMOS instrument on the ESO Very Large Telescope \citep{LeFevre2003}. VIPERS provides high-quality spectra for over 86\,000 galaxies in the redshift range $0.5 < z < 1.2$, selected to a magnitude limit of $i_{\mathrm{AB}} \leq 22.5$. The survey covers approximately 23.5 deg$^2$ across the CFHTLS-W1 (15.7 deg$^2$) and CFHTLS-W4 (7.8 deg$^2$) fields.

The signal-to-noise ratio (S/N) of the spectra depends primarily on galaxy brightness and observing conditions, and is quantified through a redshift quality flag (z$\_\rm{flag}$). High-confidence redshift measurements correspond to spectra with strong features and S/N sufficient for secure redshift determination (typically $>90\%$ confidence). Only galaxies with z$\_\rm{flag}$ values between 2 and 9, indicating reliable redshift estimates, are included in the analysis \citep{Guzzo2014, Garilli2014, Scodeggio2016}.

To construct their sample of dwarf galaxies, \cite{Siudek2023a} selected 33\,333 VIPERS galaxies with stellar masses below $\log(M_\star/M_\odot) = 10.0$ and reliable redshift measurements ($\geq 90\%$ confidence). This stellar mass threshold is slightly more inclusive than the commonly adopted value for dwarf galaxies (typically $\log(M_\star/M_\odot) \lesssim 9.5$), but it is consistent with other studies focused on AGN in low-mass systems and allows for a statistically significant sample across environments. Importantly, a high fraction of galaxies in the $9<\log(M_\star/M_\odot) < 10$ range are expected to host intermediate-mass black holes. After applying various quality criteria their final sample consisted of 12\,942 dwarf galaxies in the redshift range $\rm 0.5 < z \leq 0.9$. 

\begin{figure*}
\centering
   \includegraphics[width=2.\columnwidth, height=7.5cm]{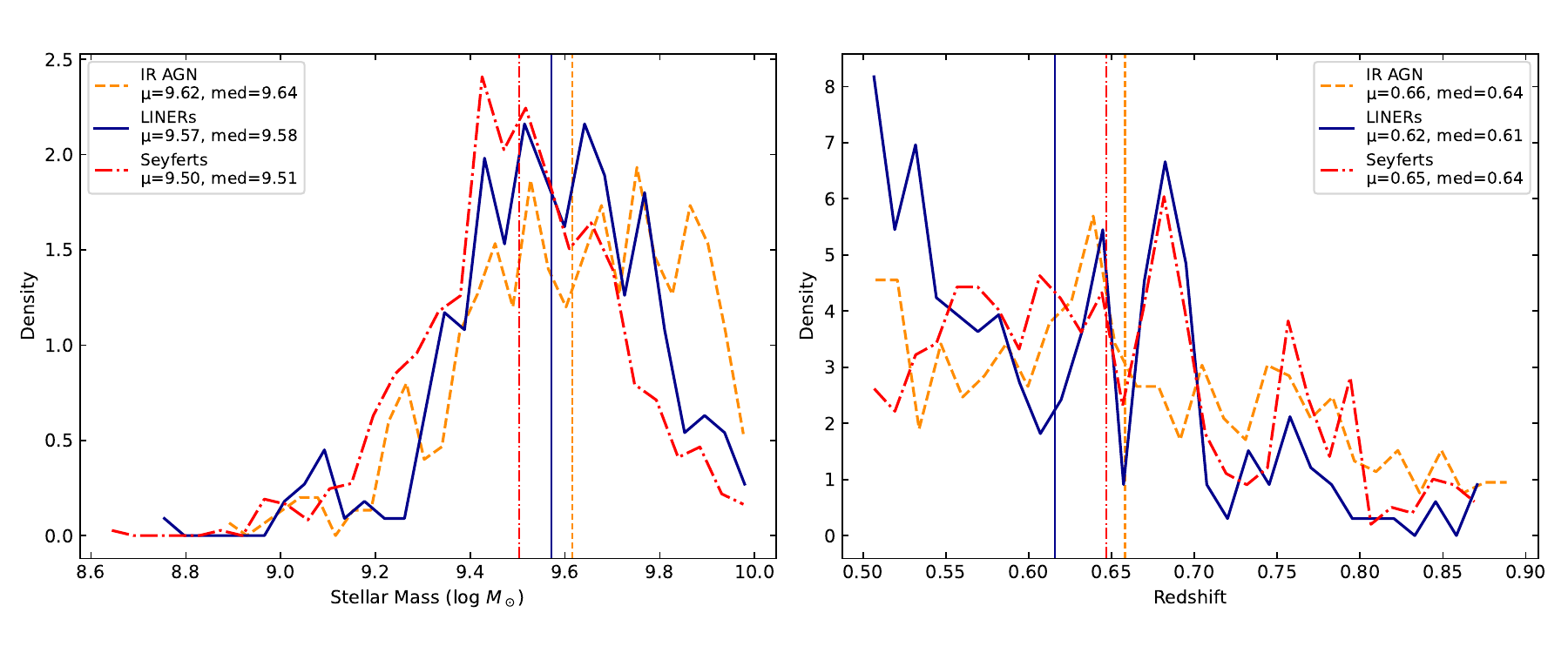}   
  \caption{Distributions of stellar mass (left) and redshift (right) for the three AGN populations. The mean ($\mu$) and median (med) values for each distribution are indicated in the legends. Vertical lines correspond to mean values.}
  \label{fig_agn_distrib}
\end{figure*} 

\begin{figure}
\centering
   \includegraphics[width=0.9\columnwidth, height=6.5cm]{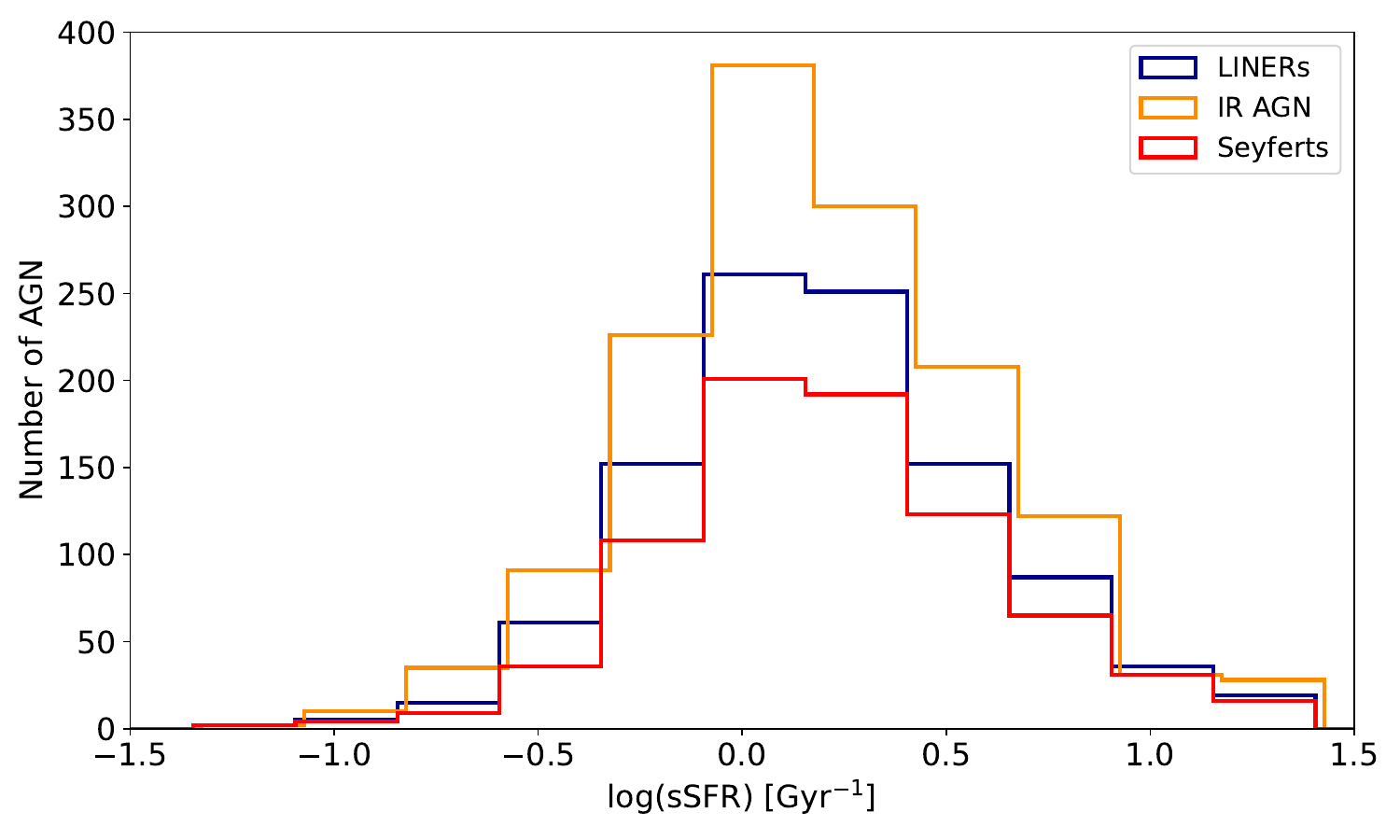}   
  \caption{Distributions of specific SFR ($\rm sSFR=\frac{SFR}{M\star}$) for the three AGN populations.}
  \label{fig_ssfr}
\end{figure}

\subsection{The AGN datasets}
\label{sec_agn}

To identify AGN within the final sample of 12\,942 dwarf galaxies, \cite{Siudek2023a} employed two complementary selection methods. The first is based on an emission-line diagnostic diagram using [OII]$\lambda$3726, H$\beta$, and [OIII]$\lambda$5007, following the methodology of \cite{Lamareille2010}. This approach extends traditional BPT diagnostics \citep{Baldwin1981} to higher redshifts ($\rm z > 0.45$), where H$\alpha$ and NII are no longer within the observed spectral range. Emission line equivalent widths (EWs) rather than fluxes were used to mitigate the effects of dust attenuation. After applying a series of quality criteria to ensure reliable line detections—including constraints on line width, peak position, and amplitude, a final sample of 4\,315 dwarf galaxies with high-confidence line measurements was selected. From these, 1\,050 were classified as AGN (787 Seyferts and 263 LINERs) based on their position in the diagnostic diagram.

A second, independent AGN sample was identified using mid-infrared photometry from WISE, matched to the optical dataset within a 10-arcsecond radius. AGN candidates were selected based on mid-infrared color cuts proposed by \cite{Hviding2022}, optimized for high completeness. This yielded 393 additional infrared (IR) AGN not overlapping with the optical sample. We note that M$_\star$ are derived from full SED fitting across UV to near-IR wavelengths, which naturally accounts for dust extinction via reddening and template fitting \citep[see section \ref{sec_physical_prop} and ][]{Moutard2016}. Additionally, SFRs are corrected for extinction using the Balmer decrement where available, or derived from [OII] luminosities following established prescriptions that include dust corrections \citep[e.g.,][]{Kewley2004}. Thus, dust attenuation has been consistently treated in the estimation of host galaxy properties for all AGN types, including the IR-selected sample. Fig. \ref{fig_agn_distrib} presents the M$_\star$ and redshift distributions of the three AGN datasets. The three AGN populations exhibit similar distributions in both M$_\star$ and redshift. Kolmogorov–Smirnov (KS) test yield a p-value higher than 0.6 in all cases. 

\begin{figure*}
\centering
   \includegraphics[width=1.\columnwidth, height=6.8cm]{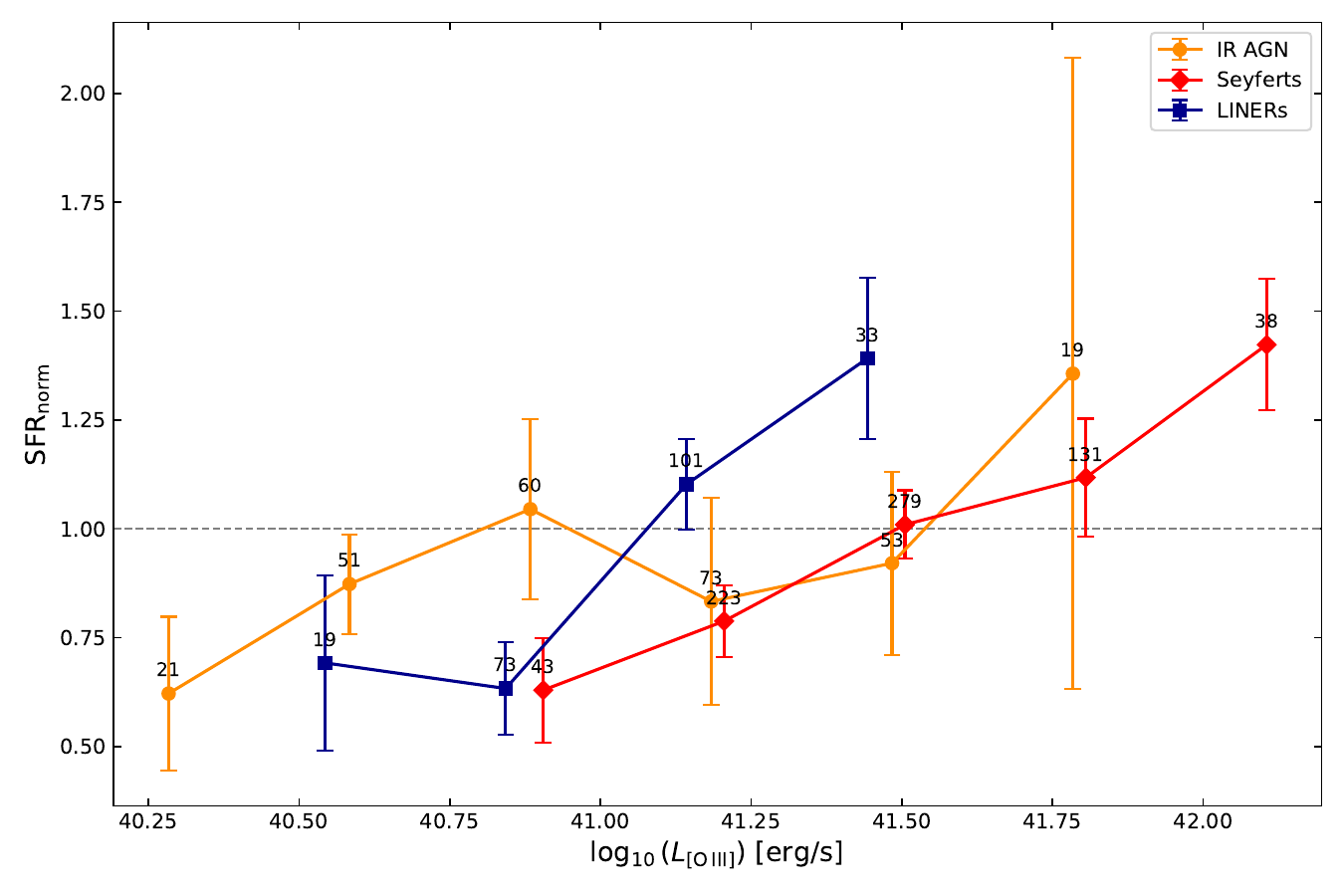}  
   \includegraphics[width=1.\columnwidth, height=6.8cm]{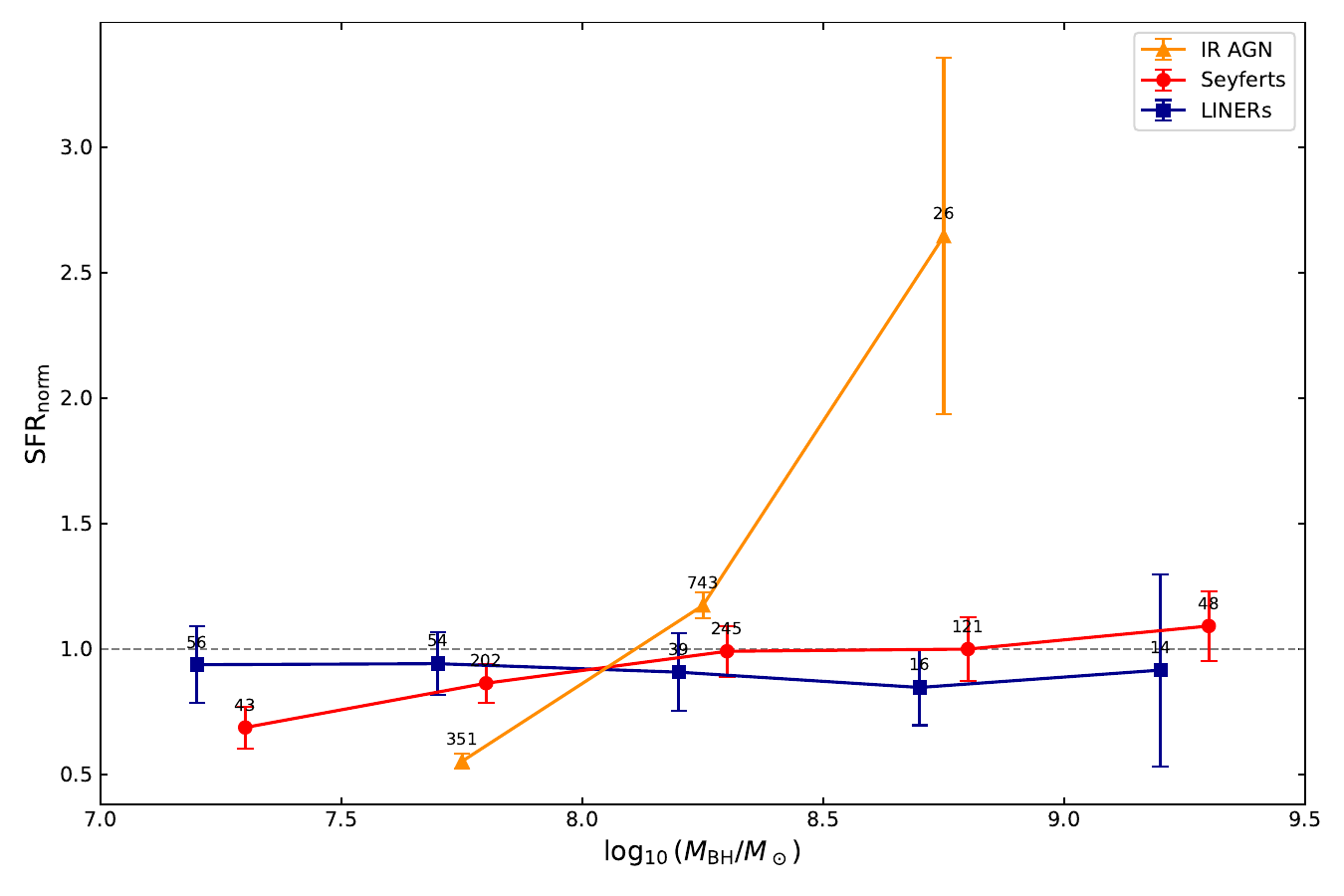}  
   \includegraphics[width=1.\columnwidth, height=6.8cm]{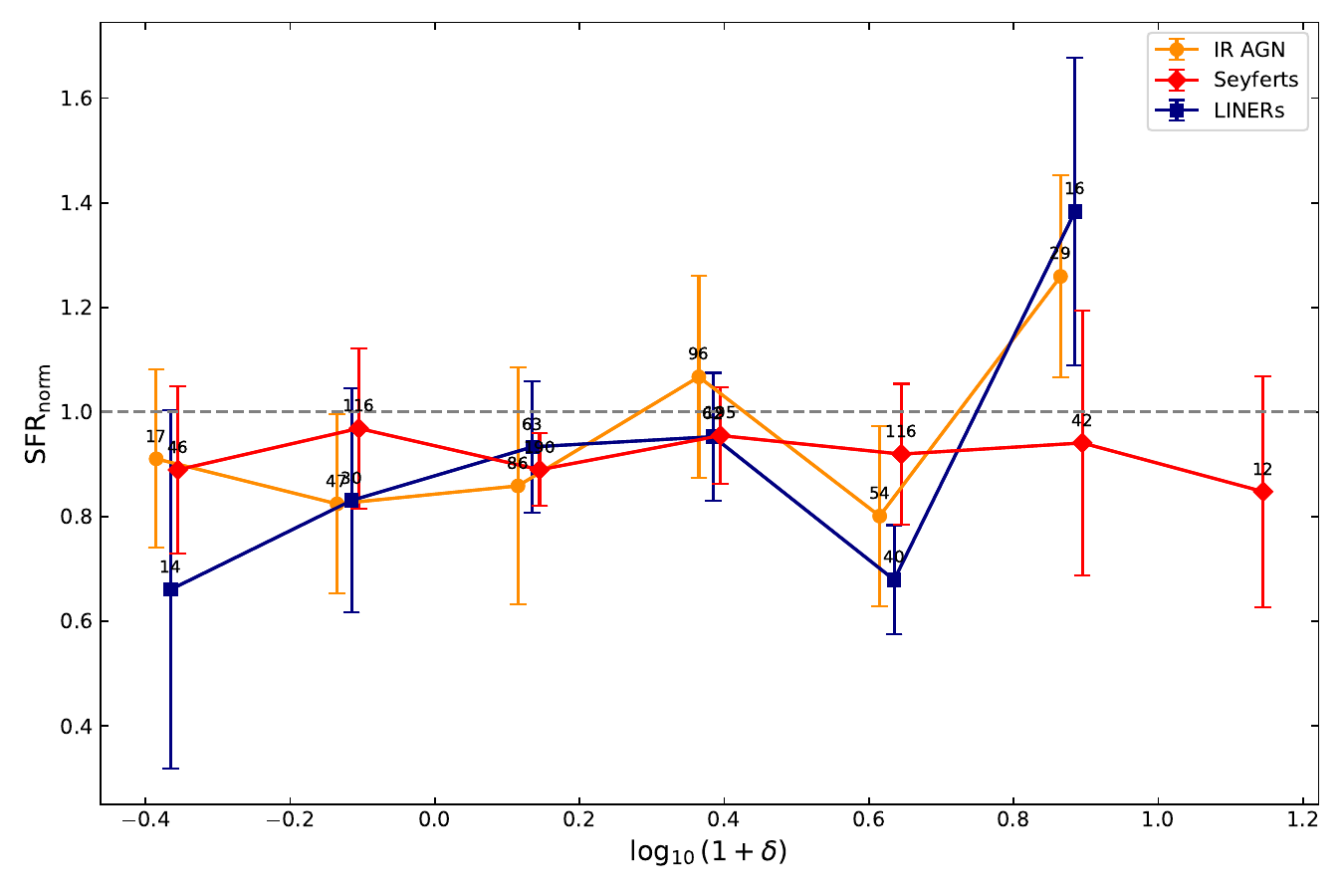}  
   \includegraphics[width=1.\columnwidth, height=6.8cm]{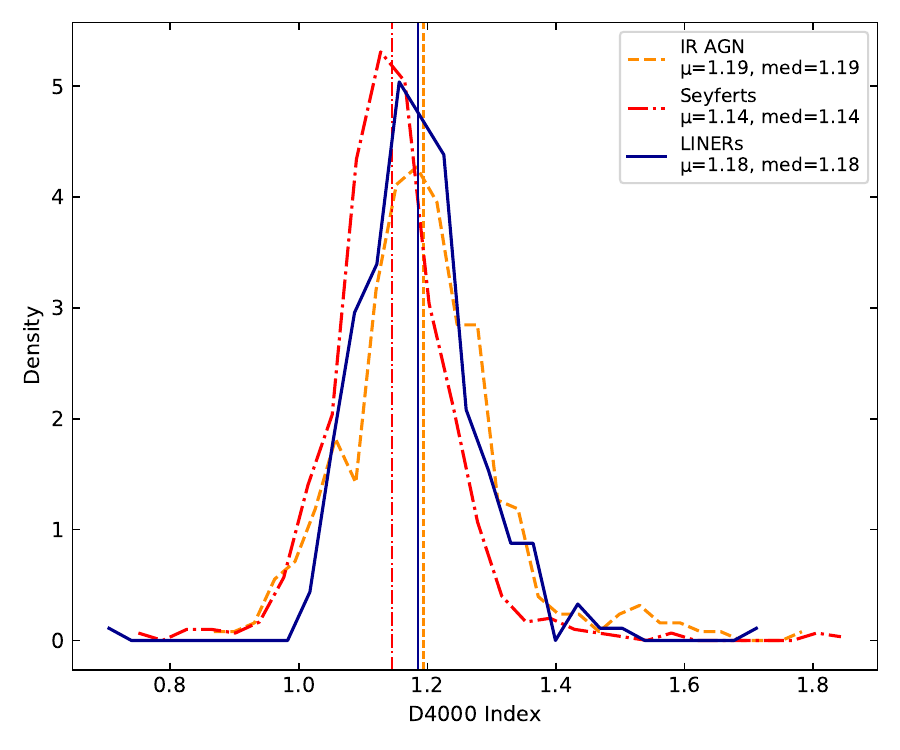}  
  \caption{Normalized star formation rate (SFR$_\mathrm{norm}$) for the three AGN populations as a function of key physical parameters. The top left panel shows SFR$_\mathrm{norm}$ versus [OIII] luminosity, the top right panel shows SFR$_\mathrm{norm}$ as a function of black hole mass, and the bottom left panel presents SFR$_\mathrm{norm}$ as a function of environmental density. The bottom right panel displays the distributions of stellar ages, using the D4000 index as a proxy.}
  \label{fig_sfrnorm}
\end{figure*} 

\subsection{Control sample}
\label{sec_control}

To enable a fair comparison with AGN hosts, a control sample of 1\,050 star-forming galaxies was constructed by matching each AGN to a non-AGN galaxy with similar stellar mass, redshift, and optical color (r‒i). The matching procedure follows the methodology of \cite{Cheung2015} and \cite{Kristensen2020}, using a nearest-neighbor approach in three-dimensional parameter space.  As a result, the control sample closely mirrors the AGN population in terms of key physical properties. The average differences in stellar mass and color between the AGN and control samples are less than 0.2 times the standard deviation of the corresponding distribution for the AGN sample. Statistical tests confirm that the two samples are indistinguishable in stellar mass and color across all redshift bins \citep[][]{Siudek2023a}.

\subsection{Physical properties}
\label{sec_physical_prop}

A large number of physical properties have been calculated in \cite{Siudek2023a} \citep[see also][]{Moutard2016} and are available. M$_\star$ has been derived  through SED fitting using the LePhare code \citep{Ilbert2006} and a library of templates based on \cite{Bruzual_Charlot2003} models with a \cite{Chabrier2003} initial mass function (IMF). Photometric inputs span UV to near-IR wavelengths. The SFR  is derived from extinction-corrected [OII] luminosity following \cite{Kewley2004}, which provides better consistency across the redshift range given limited far-IR coverage.

Galaxy structural parameters (effective radius $R_e$ and S\'ersic index $n$) were obtained from GALFIT modeling \cite{Peng2002} of CFHTLS $i$-band images \citep{Krywult2017}. Environmental densities are characterized using the local overdensity parameter $\delta$ from \cite{Cucciati2017}, computed within cylinders defined by the 5th-nearest neighbor approach and calibrated to avoid redshift evolution biases. 

Black hole masses for the optical AGN in our sample have also been measured in \citet{Siudek2023a}. For their calculation the empirical calibration from \citet{Baron2019} was combined with Equation 1 of \citet{FerreMateu2021}. The method is designed for use with narrow-line AGN and does not require the presence of broad emission lines. It is based on an empirical correlation between the narrow-line luminosity ratio $L(\mathrm{[O\,III]})/L(\mathrm{H}\beta)$ and the full width at half maximum (FWHM) of the broad H$\alpha$ emission line.
 
Specifically, black hole masses were estimated using the following equation:

\begin{equation}
\log \left( \frac{M_{\rm BH}}{M_\odot} \right) = 3.55 \log \left( \frac{L_{\rm [OIII]}}{L_{\rm H\beta}} \right) + 0.59 \log L_{\rm bol} - 20.96,
\end{equation}
as proposed by \citet{Baron2019}, where $L_{\rm [OIII]}$ and $L_{\rm H\beta}$ are the extinction-corrected luminosities of the [O\,\textsc{iii}]$\lambda$5007 and H$\beta$ lines. The bolometric luminosity $L_{\rm bol}$ is estimated following the prescription in \citet{Netzer2009}:

\begin{equation}
\log L_{\rm bol} = \log L_{\rm H\beta} + 3.48 + \max \left[ 0,\, 0.31 \left( \log \left( \frac{L_{\rm [OIII]}}{L_{\rm H\beta}} \right) - 0.6 \right) \right].
\end{equation}
This formalism has been shown to correlate with black hole masses derived via stellar velocity dispersions or broad-line virial methods, particularly for Type 2 AGN. The specific combination of emission line ratios and bolometric corrections in Equation 1 of \citet{FerreMateu2021} provides a practical calibration that can be applied to large AGN samples lacking BLR measurements.

While the method introduces a non-negligible scatter (up to $\sim1$ dex; \citealt{Reis2020}), it does not appear to introduce a systematic offset in $M_{\rm BH}$. The technique has been employed successfully in previous studies \citep[e.g.][]{Vietri2022}, including low-mass AGN and dwarf galaxies \citep[e.g.,][]{Siudek2023a}. Given the large number of sources in our sample and our use of binned statistics throughout the analysis (see sections \ref{sec_results} and \ref{sec_discussion}), the effects of this intrinsic scatter should be diluted when interpreting global trends. We nevertheless caution that individual $M_{\rm BH}$ estimates may carry large uncertainties.

\subsection{Calculation of SFR$_{norm}$}
\label{sec_sfrnorm}

To investigate the star formation properties of different AGN types relative to star-forming dwarf galaxies, we use the SFR$_\text{norm}$ parameter \citep[e.g.,][]{Mullaney2015, Masoura2018, Mountrichas2021c, Mountrichas2022a, Mountrichas2022b, Koutoulidis2022, Pouliasis2022}. This parameter is defined as the ratio of the SFR of an AGN host to that of non-AGN star-forming galaxies with similar properties. Specifically, for each AGN, we identify all galaxies from the control sample described in Section~\ref{sec_control} that lie within $\pm$0.1 dex in stellar mass and $\pm$0.1 in redshift. We then compute the SFR ratio between the AGN and each of these matched galaxies. The median of these individual ratios is adopted as the SFR$_\mathrm{norm}$ of the AGN. Uncertainties are estimated via bootstrap resampling and correspond to the 1$\sigma$ scatter. As shown in \citet{Mountrichas2021c}, this approach is robust to the exact choice of matching window, although narrower ranges may yield fewer matches. We restrict our analysis to AGN with at least 50 matched galaxies in the control sample. Additionally, only bins containing a minimum of 10 AGN are shown throughout the analysis to ensure statistical robustness.

To compute the SFR$_\mathrm{norm}$ parameter, it is essential to identify and exclude quiescent systems from both the AGN and non-AGN galaxy samples. The galaxy control sample already includes only star-forming galaxies (Section~\ref{sec_control}). To identify quiescent systems within the AGN datasets, we examine their specific star formation rates (sSFR). Figure~\ref{fig_ssfr} presents the $\log(\mathrm{sSFR})\ [\mathrm{Gyr}^{-1}]$ distributions for the AGN populations used in this work. All three AGN samples exhibit a prominent peak at similar values (around 0.03), indicating no evidence of a secondary, lower peak that would typically signify a quiescent sub-population \citep[e.g.,][]{Mountrichas2021c, Mountrichas2022a, Mountrichas2022b}.

Alternative criteria to define quiescent systems have also been explored in the literature. For example, \citet{Salim2018} adopt a threshold of 1 dex below the mean sSFR, yet only a very small number of AGN in our samples fall below this limit (see Fig.~\ref{fig_ssfr}). Similarly, studies such as \citet{Franx2008} have employed a fixed cut at $\log(\mathrm{sSFR})\ [\mathrm{Gyr}^{-1}] = -2$, but none of the AGN in our datasets satisfy this condition.

A more conservative definition, where quiescent systems are those with $\log(\mathrm{sSFR})$ values at least 0.3 dex below the mean \citep[e.g.,][]{Muzzin2013a, Shimizu2015, Koutoulidis2022}, identifies roughly 9–12\% of AGN as quiescent in our sample. Despite this, we choose to retain these sources in our analysis. We have verified that excluding them does not impact our main conclusions: while the SFR$_\mathrm{norm}$ values would increase slightly (by approximately 0.1 dex), this shift lies well within the statistical uncertainties of our measurements.

\begin{figure}
\centering
  \includegraphics[width=0.99\columnwidth, height=7.5cm]{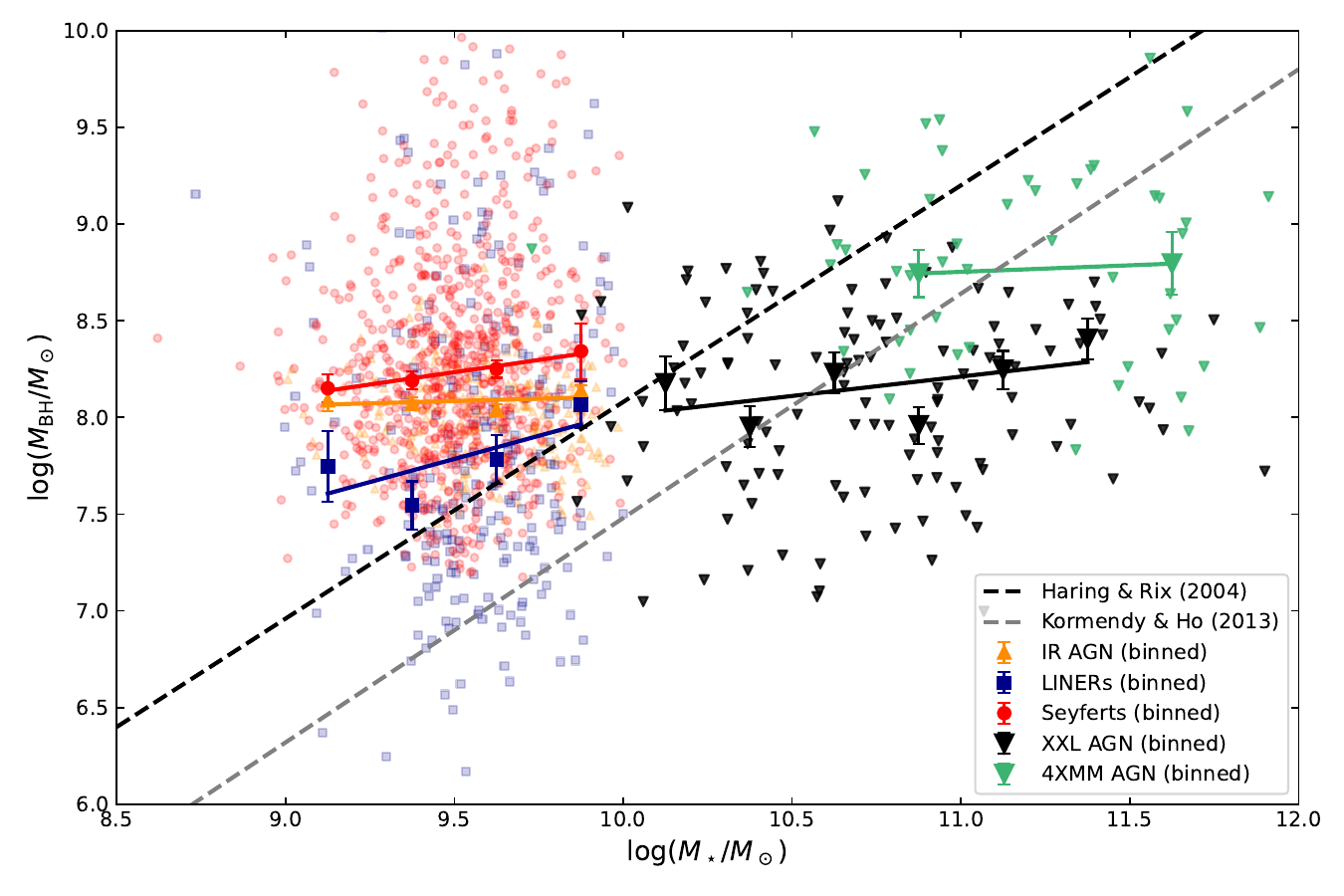}   
  \caption{Black hole mass versus stellar mass for the three AGN populations. Small colored symbols represent individual sources, with distinct shapes and colors indicating the different AGN populations as shown in the legend. Large symbols indicate median values in stellar mass bins of width 0.25\,dex. Error bars correspond to the standard error of the mean ($1\,\sigma$) in each stellar mass bin, computed as the standard deviation divided by the square root of the number of sources in the bin. Solid lines show least-squares fits to the binned data. Dashed lines correspond to established $M_\mathrm{BH}$–$M_\star$ relations in the local universe from the literature. Black and green markers denote X-ray selected AGN from the XMM-XXL and 4XMM surveys, respectively, restricted to the same redshift range as our sample ($\rm 0.5 < z < 0.9$). Corresponding solid lines show the best-fit relations for these X-ray AGN samples.}
  \label{fig_mbh_mstar}
\end{figure}

\section{Results}
\label{sec_results}

In this section, we present a comparative analysis of the star formation rates (SFRs) of different AGN populations and star-forming galaxies, exploring their dependence on AGN power, black hole mass, and cosmic environment. We also examine differences in their stellar populations and investigate the M$_\mathrm{BH}$–M$_\star$ relation across the various AGN types.

\subsection{AGN and star-forming dwarfs: star formation across luminosity, black hole mass and environment}

The top left panel of Figure~\ref{fig_sfrnorm} presents SFR$_\text{norm}$ as a function of $\log L\text{[OIII]}$ for the various AGN populations. All AGN types show similar trends, regardless of selection method: at low luminosities, AGN in dwarf galaxies exhibit suppressed star formation compared to star-forming galaxies of similar mass (SFR$_\text{norm} < 1$). As luminosity increases, SFR$_\text{norm}$ rises, and above a threshold of log\,L[OIII], AGN begin to show enhanced star formation relative to the control sample.

These trends mirror those observed for X-ray selected AGN in more massive galaxies ($10 < \log (M_\odot) < 12$), where AGN generally display lower or comparable SFRs to star-forming galaxies at low and intermediate X-ray luminosities (L$_X$), with enhanced SFR only emerging above a luminosity threshold \citep[][]{Masoura2018, Pouliasis2022}. Notably, the L$_X$ threshold for this transition increases with the stellar mass of the host galaxy \citep[][]{Mountrichas2021c, Mountrichas2022a, Mountrichas2022b, Mountrichas2024c, Cristello2024}.

While all three AGN types exhibit a general increase in SFR$_\text{norm}$ with luminosity, the luminosity threshold at which this transition occurs differs across populations. LINERs show enhanced star formation above $\log L\mathrm{[OIII]} \sim 41$, Seyferts at $\sim 41.75$, and IR AGN fall in between, though with larger associated uncertainties. The observed variation in the luminosity threshold at which SFR$_\text{norm}$ exceeds unity among AGN types may reflect intrinsic differences in their fueling mechanisms or evolutionary stage. It is important to note, though, that for IR AGN, which are often heavily obscured, the observed [OIII] luminosity likely underestimates the intrinsic AGN power due to dust extinction. As a result, their SFR$_\mathrm{norm}$ values at a given L[OIII] should be interpreted with caution, as these AGN may be intrinsically more powerful than suggested by their optical line luminosity.

The top right panel of Fig.~\ref{fig_sfrnorm} shows SFR$_\text{norm}$ as a function of black hole mass. LINERs exhibit an almost flat trend, with SFR$_\text{norm}$ values systematically below unity, though largely consistent with it within uncertainties. Seyferts, in contrast, display a mild upward trend, suggesting a moderate increase in SFR$_\text{norm}$ with black hole mass. IR-selected AGN demonstrate a more pronounced rise in SFR$_\text{norm}$ as M$_{BH}$ increases and also a narrower range of black hole masses compared to LINERS and Seyferts. A similar trend, that is SFR enhancement with increasing black hole mass, has also been observed in X-ray selected AGN \citep{Mountrichas2023d}. It is also worth noting that in both IR-selected and X-ray-detected AGN, SFR$_\text{norm}$ exceeds unity at comparable black hole masses.

The bottom left panel of Figure~\ref{fig_sfrnorm} shows the SFR$_\mathrm{norm}$ as a function of environment density, $\log\,(1+\delta)$, for the different AGN populations. Seyferts maintain relatively flat SFR$_\mathrm{norm}$ values below unity across all environments. LINERs and IR AGN, on the other hand, show a possible upturn in SFR$_\mathrm{norm}$ at the densest environments, though this is driven by a single bin and should be interpreted with caution.

Finally, the bottom right panel of Figure~\ref{fig_sfrnorm} shows the distribution of the D4000 index for the three AGN populations. While the mean and median values across IR AGN, Seyferts, and LINERs differ only modestly, KS tests indicate that these differences are statistically significant (p-value < 0.01 in all cases). These results suggest that Seyfert hosts tend to exhibit younger stellar populations compared to LINERs and IR AGN. We note that D4000 measurements in VIPERS have typical uncertainties of $\lesssim 0.1$ \citep{Garilli2014}. Although these uncertainties are not propagated into the KS tests, the large sample sizes and modest typical errors ensure that the statistical significance of the distribution differences remains robust.

\subsection{The M$_{BH}-$M$_\star$ relation of different AGN populations in dwarf galaxies}

In this section, we examine the relationship between M$_\mathrm{BH}$ and M$_\star$ and the $\log(M_{\mathrm{BH}} / M_\star)$ ratio as a function of redshift and M$_\star$ for different AGN populations hosted in dwarf galaxies. Fig.~\ref{fig_mbh_mstar} present the M$_\mathrm{BH}$ versus M$_\star$. Small symbols represent individual measurements, while larger markers show the median black hole masses within stellar mass bins of width 0.25\,dex. Error bars correspond to the standard error of the mean ($1\,\sigma$) in each stellar mass bin, computed as the standard deviation divided by the square root of the number of sources in the bin. Solid lines indicate least-squares linear fits to the binned data. We find that fits to the individual sources yield consistent results, and a similarly good agreement is obtained using the more robust Theil-Sen fitting method. The best-fitting relations (least squares on binned data) for each AGN population are:
\begin{eqnarray*}
\text{IR AGN } &:& \log(M_\mathrm{BH}/M_\odot) = (0.048 \pm 0.019)\log(M_\star/M_\odot) \\
               & & \hspace{3.7cm} + (7.664 \pm 0.804) \\
\text{LINERs } &:& \log(M_\mathrm{BH}/M_\odot) = (0.478 \pm 0.248)\log(M_\star/M_\odot) \\
               & & \hspace{3.7cm} + (3.247 \pm 2.115) \\
\text{Seyferts } &:& \log(M_\mathrm{BH}/M_\odot) = (0.254 \pm 0.033)\log(M_\star/M_\odot) \\
               & & \hspace{3.7cm} + (5.821 \pm 0.311)
\end{eqnarray*}
The quoted uncertainties on the slope and intercept reflect the standard error of the linear regression (computed using \texttt{scipy.stats.linregress}). These values indicate the robustness of the trends for each AGN type: the relatively small errors for Seyferts and IR AGN suggest well-constrained correlations, while the larger uncertainties for LINERs reflect their more scattered distribution in the M$_{\mathrm{BH}}$--M$_\star$ plane, possibly due to intrinsic diversity or measurement limitations.

We observe that, at fixed M$_\star$, Seyfert galaxies tend to host more massive black holes than LINERs. This indicates that Seyferts may lie on a different M$_\mathrm{BH}$–M$_\star$ scaling relation, potentially reflecting earlier or more rapid SMBH growth histories. In contrast, LINERs exhibit a steeper M$_\mathrm{BH}$–M$_\star$ relation, indicating a stronger coupling between black hole and stellar mass in this population. This trend is reminiscent of the steeper local scaling relations observed in bulge-dominated or quenched galaxies \citep[e.g.,][dashed lines]{Haring2004, Kormendy2013}. IR-selected AGN display a flatter relation overall, aligning more closely with Seyferts in normalization but showing little dependence on stellar mass, possibly reflecting a different fueling mode or evolutionary stage.

\section{Discussion}
\label{sec_discussion}

The AGN populations examined in this work are selected using distinct diagnostics, each sensitive to different aspects of black hole accretion and host galaxy conditions. Seyferts, characterized by strong high-ionization lines, are generally associated with radiatively efficient accretion onto SMBHs, likely fueled by cold gas inflows or internal dynamical processes such as bars and minor mergers \citep[e.g.,][]{Kewley2006, Ellison2011, Heckman2014}. In contrast, LINERs exhibit low-ionization emission-line spectra, consistent with radiatively inefficient accretion \citep{Ho2008} or alternative ionizing sources such as shocks or post-AGB stars \citep[e.g.,][]{Heckman1989, Sarzi2010, Singh2013, Belfiore2016}. Their host galaxies tend to be older, more massive, and more passive in terms of star formation \citep{CidFernandes2011, Gavazzi2018}. IR-selected AGN, identified via WISE mid-infrared colors, trace hot dust emission from obscured accretion activity \citep{Stern2012, Assef2013}. This selection is particularly sensitive to heavily obscured and possibly Compton-thick AGN phases \citep{Donley2012, Hickox2018}.

Our comparison of SFR${_\mathrm{norm}}$ as a function of AGN power, traced by [OIII] luminosity, reveals several systematic trends (top, left panel of Fig. \ref{fig_sfrnorm}). All three AGN populations show suppressed star formation relative to non-AGN star-forming galaxies at low L${\mathrm{[OIII]}}$, with SFR${_\mathrm{norm}} < 1$. As L${\mathrm{[OIII]}}$ increases, SFR$_{\mathrm{norm}}$ rises and eventually exceeds unity, albeit at different luminosity thresholds. These thresholds likely reflect differences in accretion modes, dust obscuration, selection biases, and potentially evolutionary stages.

The low L${\mathrm{[OIII]}}$ values in IR AGN may underrepresent their true accretion power, as the optical [OIII] line is more affected by dust extinction than mid-IR emission. Therefore, the observed [OIII] luminosity in IR AGN may significantly underestimate the intrinsic AGN power. As a result, the luminosity threshold at which SFR${_\mathrm{norm}}$ begins to rise may appear at lower observed L${\mathrm{[OIII]}}$ than its true, intrinsic value. The IR AGN population may represent an early or dust-enshrouded AGN phase, where SMBH accretion and star formation co-occur in a gas-rich environment, consistent with the high SFR${_\mathrm{norm}}$ observed at high L${\mathrm{[OIII]}}$.

In contrast, LINERs likely include a mixture of weakly accreting AGN and non-AGN sources powered by shocks or old stellar populations, leading to their overall lower SFR${_\mathrm{norm}}$. The shallow increase in SFR${_\mathrm{norm}}$ at low luminosities may be driven by a small subset of true AGN in the LINER population that become detectable near $\log {\mathrm{L[OIII]}} \sim 41$.

Examining SFR$_{\mathrm{norm}}$ as a function of black hole mass (top, right panel of Fig. \ref{fig_sfrnorm}), we find that LINERs display values broadly consistent with unity across the entire M$_{\mathrm{BH}}$ range, suggesting no strong dependence of star formation on black hole mass in this population. Seyferts show a modest increase in SFR$_{\mathrm{norm}}$ with M$_{\mathrm{BH}}$, although values remain near unity, particularly at $\log(M_{\mathrm{BH}}/M_\odot) > 8$. IR AGN exhibit a more pronounced positive trend, with higher M$_{\mathrm{BH}}$ associated with enhanced SFR$_{\mathrm{norm}}$, potentially reflecting greater gas availability or more efficient star formation in dust-obscured systems. These results align with previous findings for X-ray selected AGN, which also report a weak but positive correlation between star formation and black hole mass \citep{Mountrichas2023d}.

Environmental density can influence gas availability and galaxy interactions. In our sample, SFR$_{\mathrm{norm}}$ shows no strong or systematic dependence on local overdensity, with LINERs, Seyferts, and IR AGN exhibiting broadly similar trends across the range of $\log(1 + \delta)$ (bottom, left panel of Fig. \ref{fig_sfrnorm}). If any differences exist, they are subtle and could be masked by scatter or selection effects. This suggests that, within the observed density range, the large-scale environment may play a secondary role compared to internal galaxy properties or AGN type in regulating star formation.

We also examined stellar population ages, as traced by the D4000 index (bottom, right panel of Fig. \ref{fig_sfrnorm}). LINERs and IR AGN host older stellar populations (median D4000 $\approx$ 1.18–1.19), while Seyfert hosts are slightly younger (D4000 $\approx$ 1.14). A Kolmogorov–Smirnov test confirms these differences are statistically significant ($p < 0.01$ in all cases), supporting a picture in which LINERs are more evolved and possibly in a post-starburst phase.

Our analysis of the M${_\mathrm{BH}}$–M$_\star$ relation and the evolution of the $\log(M_{\mathrm{BH}} / M_\star)$ ratio across different AGN populations offers insights into the co-evolution of black holes and their host galaxies in low-mass systems. The diversity in scaling relations among LINERs, Seyferts, and IR-selected AGN suggests that the nature and efficiency of black hole growth are closely linked to both the AGN accretion mode and host galaxy conditions. 

Seyferts show elevated black hole masses at fixed stellar mass compared to LINERs, suggesting that they may lie on a different M$_{\mathrm{BH}}$–M$_\star$ scaling relation (Fig. \ref{fig_mbh_mstar}). This offset could reflect earlier black hole growth episodes or differences in black hole fueling efficiency. The interpretation is consistent with their younger stellar populations and higher SFR$_{\mathrm{norm}}$. The relatively flatter M$_{\mathrm{BH}}$–M$_\star$ relation for Seyferts and IR AGN may indicate a decoupling between black hole and stellar mass growth at low masses, potentially driven by intermittent fueling or bursty star formation. In contrast, the steeper relation observed for LINERs implies a tighter coupling between M$_{\mathrm{BH}}$ and M$_\star$, suggestive of co-evolution under more stable or feedback-regulated conditions. These results are in line with local observations showing that quenched, bulge-dominated galaxies tend to follow steeper M$_{\mathrm{BH}}$–M$_\star$ relations \citep[e.g.,][]{Haring2004, Kormendy2013}.

We should note though that the $M_{\mathrm{BH}}$ for the sources used in our analysis are estimated using narrow emission lines and bolometric corrections (see Sect. \ref{sec_physical_prop}). While this approach allows $M_{\mathrm{BH}}$ estimates in the absence of broad lines, it is inherently less direct than virial methods and subject to additional uncertainties. Specifically, it relies on extinction corrections, assumptions about bolometric luminosities, and the calibration of empirical scaling relations, all of which can introduce scatter. These factors may contribute to the observed dispersion in the M$_{\mathrm{BH}}$–M$_\star$ relation, particularly for lower-mass systems where the underlying scaling may already be more uncertain.

\begin{figure}
\centering
  \includegraphics[width=1.\columnwidth, height=8cm]{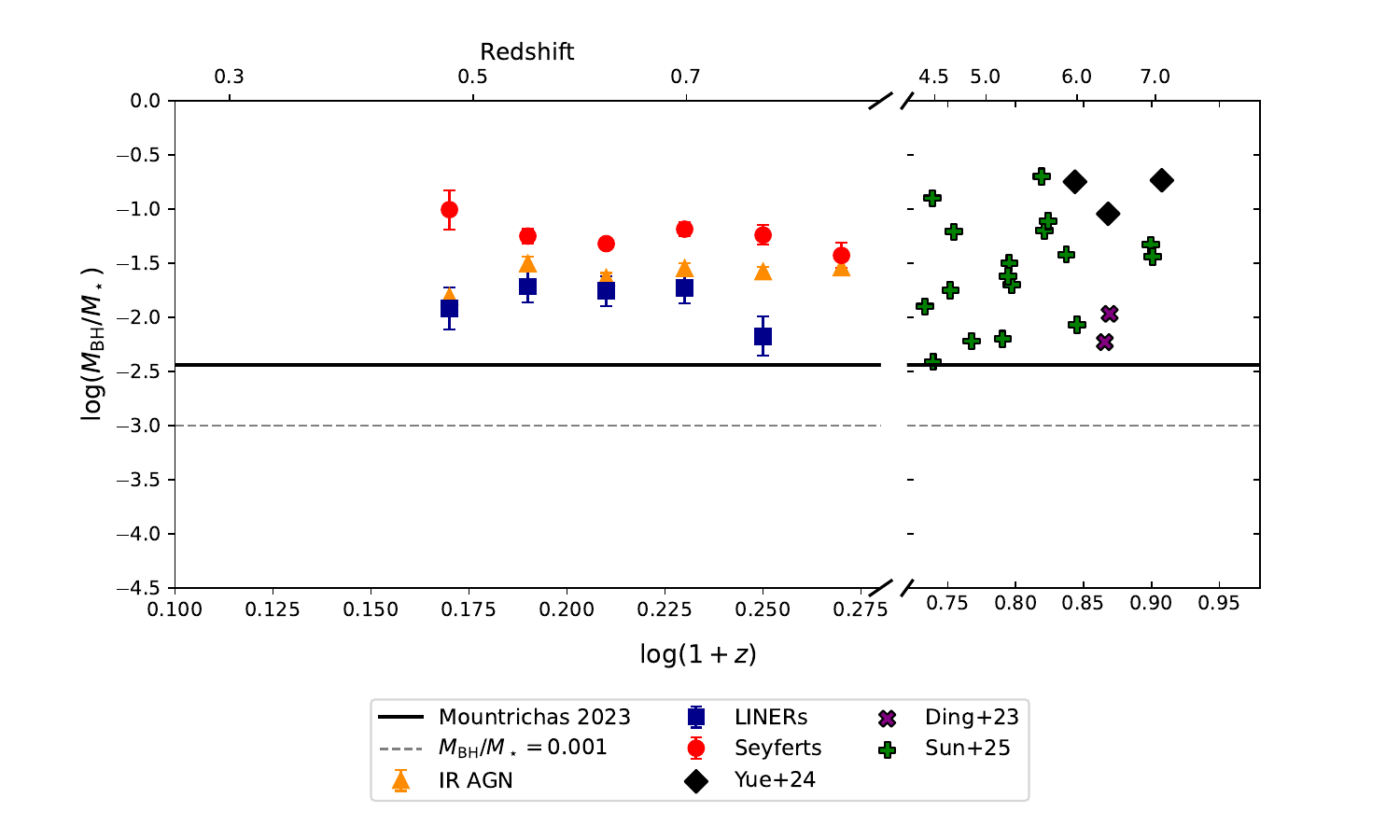}   
  \includegraphics[width=1.\columnwidth, height=8cm]{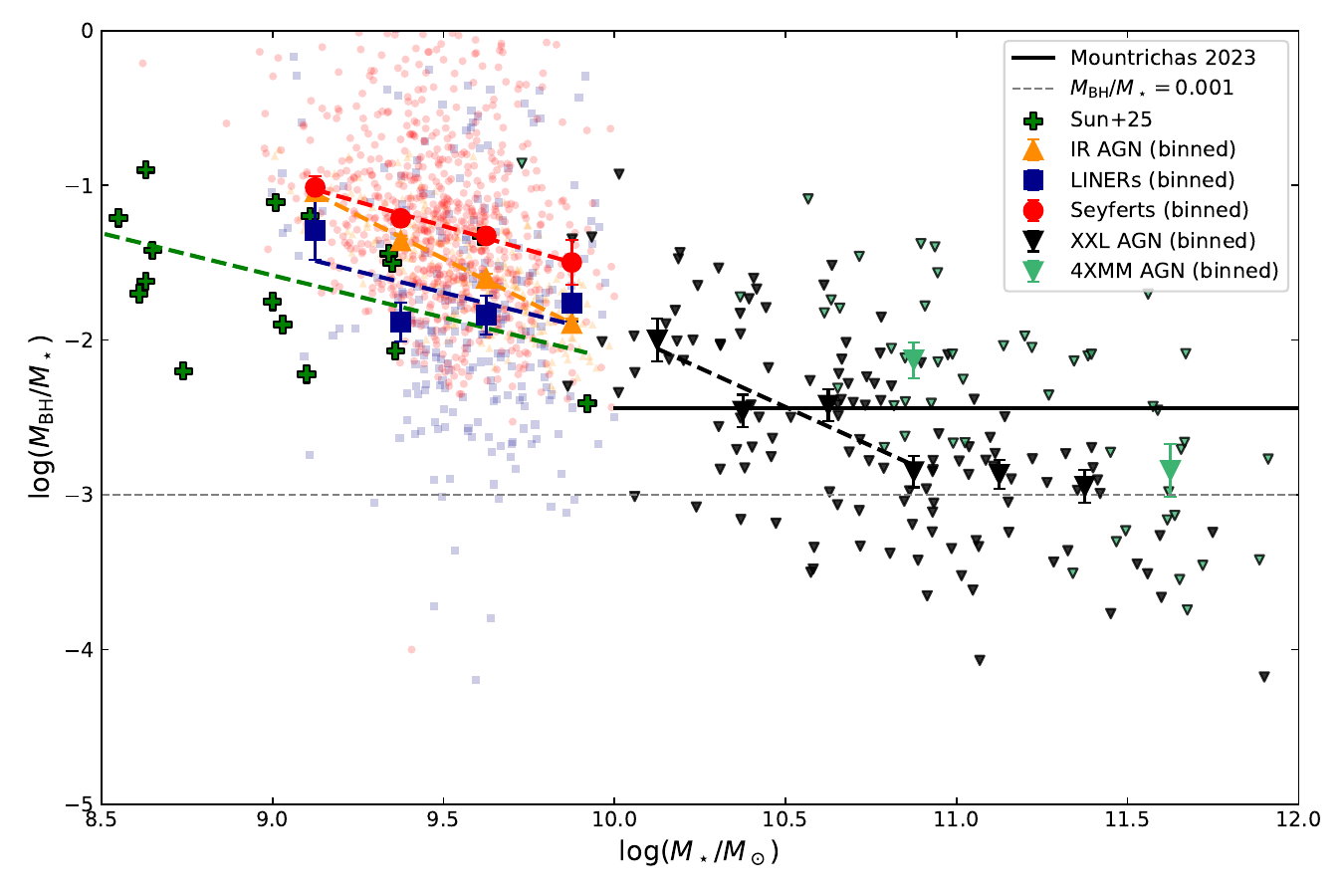} 
  \caption{Evolution of the black hole–to–stellar mass ratio.
Top: $\log(M_\mathrm{BH} / M_\star)$ vs. redshift for the three AGN populations in this study, with high-redshift ($z > 4$) AGN from the literature shown for comparison.
Bottom: $\log(M_\mathrm{BH} / M_\star)$ vs. stellar mass, including high-redshift dwarf AGN \citep{Sun2025b} and X-ray AGN from the XXL and 4XMM surveys. Dashed lines show linear fits to the binned data; for XXL AGN, the fit is limited to $\log(M_\star / M_\odot) < 11$. Solid and dashed horizontal lines indicate the average ratio for XMM-XXL AGN at $z < 2$ \citep{Mountrichas2023b} and for local inactive galaxies, respectively.
}
  \label{fig_ratio}
\end{figure} 

In addition to the dwarf galaxy AGN populations, in Fig. \ref{fig_mbh_mstar}, we also plot the M${_\mathrm{BH}}$–M$_\star$ relation for X-ray selected AGN hosted in more massive galaxies, using the 4XMM and XMM-XXL samples, presented in \cite{Mountrichas2024b} and \cite{Mountrichas2023b}. We restrict both datasets to the same redshift range as that used for the AGN populations in dwarf galaxies ($0.5 < z < 0.9$). The 4XMM sample includes 59 AGN with median $\log\,[L_{X}(\mathrm{erg,s}^{-1})] = 44.8$, for which black hole masses are derived from SDSS spectroscopy and are likely dominated by optical QSOs \citep{Wu2022}. We also include 124 X-ray detected AGN from XMM-XXL \citep{Pierre2016, Menzel2016}, with a median $\log\,[L_{X}(\mathrm{erg,s}^{-1})] = 43.7$. The best-fitting relations (least squares on binned data) for the two X-ray AGN populations are: 

\begin{eqnarray*}
\text{XXL AGN } &:& \log(M_\mathrm{BH}/M_\odot) = (0.201 \pm 0.110)\log(M_\star/M_\odot) \\
                & & \hspace{3.7cm} + (6.005 \pm 1.725) \\
\text{4XMM AGN } &:& \log(M_\mathrm{BH}/M_\odot) = (0.068 \pm 0.027)\log(M_\star/M_\odot) \\
                 & & \hspace{3.7cm} + (8.009 \pm 2.125)
\end{eqnarray*}

We find that 4XMM AGN exhibit systematically higher M$_{\mathrm{BH}}$ at fixed M$_\star$ compared to XMM-XXL AGN. This offset may be linked to differences in accretion properties. Given their higher X-ray luminosities, the 4XMM AGN may be radiating at higher Eddington ratios \citep[i.e., closer to their Eddington limit; e.g.,][]{Aird2012,Lusso2016}, which could imply either more actively accreting black holes or intrinsically more massive black holes for the same host mass. In contrast, the lower luminosities of XMM-XXL AGN may reflect a more quiescent accretion phase or a population caught at a different stage of growth. Although the two samples are selected using different criteria, they collectively help trace the M$_{\mathrm{BH}}$–M$_\star$ relation toward higher stellar masses and underscore how AGN selection effects can shape observed scaling relations.

Similar results to those of our optically selected AGN are reported by \citet{Mezcua2023}, who analyzed the M$_{\mathrm{BH}}$–M$_\star$ relation for seven broad-line AGN in dwarf galaxies from the VIPERS survey, six of which are also X-ray detected. When placed on the M$_{\mathrm{BH}}$–M$_\star$ plane, these AGN occupy a similar region to the three AGN populations examined in our study, reinforcing the consistency of our findings across different selection methods (see their Fig. 1).

In the top panel of Fig. \ref{fig_ratio}, we investigate how the $\log(M_{\mathrm{BH}} / M_\star)$ ratio evolves with redshift. Seyferts consistently exhibit higher black hole-to-stellar mass ratios compared to LINERs, primarily due to their more massive black holes. IR-selected AGN lie between these two populations. For all three AGN types, the $\log(M_{\mathrm{BH}} / M_\star)$ ratio appears roughly constant over the redshift range probed by our sample, implying that the balance between black hole and stellar mass is preserved over $0.5 < z < 0.9$ in dwarf hosts. We find median values of $\log(M_{\mathrm{BH}} / M_\star)$ equal to $-1.25$ for Seyferts, $-1.59$ for IR AGN and $-1.80$ for LINERs. Notably, these values are significantly elevated compared to those of more massive X-ray-selected AGN within the same redshift interval (i.e., $\rm 0.5<z<0.9$, with medians of $-2.43$ for 4XMM AGN and $-2.70$ for XMM-XXL AGN \citep[solid line;][]{Mountrichas2023b}, as well as compared to the typical ratio observed in local, inactive galaxies \cite[e.g.,][]{Haring2004, Kormendy2013}. This offset could indicate that dwarf AGN undergo more rapid black hole growth relative to their stellar content, consistent with the idea that black hole seeding and early growth may precede significant stellar mass buildup in low-mass systems \citep[e.g.,][]{Mountrichas2023b}.

A notable fraction of our AGN host galaxies exhibit M$_\mathrm{BH}$/M$_\star$ ratios exceeding 0.1, particularly within the LINER and Seyfert AGN populations (20-25\%). While these values may initially appear extreme, similar ratios have been reported in the literature for low-mass systems hosting actively accreting black holes. For example, \citet{Mezcua2023} identified broad-line AGN in dwarf galaxies from the VIPERS survey with M$_\mathrm{BH}$/M$_\star$ values up to and beyond 0.1 (see their Fig. 2), in line with the range observed in our study. Such overmassive black holes may result from early, rapid SMBH growth episodes or from selection effects favoring systems with relatively low stellar masses but substantial nuclear activity. Additional support for high M$_\mathrm{BH}$/M$_\star$ ratios has been provided by other studies of optically and X-ray selected AGN in the low-mass regime \citep[e.g.,][]{Reines2015}. These findings suggest that elevated M$_\mathrm{BH}$/M$_\star$ ratios in low-mass AGN hosts are not necessarily unphysical outliers but instead represent a real, albeit extreme, component of the black hole–galaxy co-evolution landscape.

To further investigate the diversity of AGN populations in dwarf galaxies, we quantified the scatter in M${_\mathrm{BH}}$ at fixed stellar mass. We computed the standard deviation of log(M$_{\mathrm{BH}}$) within $M_\star$ bins for each AGN sample. LINERs and Seyferts exhibit the largest scatter, with median values of 0.93 and 0.81 dex, respectively, significantly exceeding those measured for X-ray-selected AGN (0.51 dex for 4XMM, 0.48 dex for XXL) and for IR AGN (0.22 dex).

This enhanced scatter could reflect a combination of factors. First, narrow line based black hole mass estimates, particularly those relying on [OIII] luminosities, are subject to larger uncertainties and intrinsic scatter than broad-line methods typically used for QSOs. Second, dwarf galaxies likely exhibit greater stochasticity in their black hole fueling and growth histories, due to shallower potential wells, bursty star formation, and environmental effects. The fact that the scatter is substantially lower in the IR AGN and X-ray AGN samples, which may select more homogeneous or more actively accreting populations, supports the interpretation that part of the scatter in optical AGN reflects real physical diversity rather than solely measurement uncertainty.

In the same figure, we include high-redshift Seyfert galaxies from \citet{Sun2025b} at $z > 4$ with stellar masses in the range $8 < \log(M_\star/M_\odot) < 10$ (see their Table 1). We also show three $z \sim 6$ quasars from \citet{Yue2024}, for which reliable stellar masses are available (i.e., not upper limits): $\log(M_\star/M_\odot) = 9.81$, 10.14, and 10.64 (see their Tables 1 and 3). Additionally, we plot two quasars at $z > 6$ from \citet{Ding2023} with $\log(M_\star/M_\odot) = 10.53$ and 11.11. Interestingly, these high-redshift sources exhibit $\log(M_{\mathrm{BH}} / M_\star)$ values comparable to those of the low-mass AGN populations in our sample. Specifically, Seyferts from \citet{Sun2025b} at $z > 4$ exhibit a median $\log(M_{\mathrm{BH}} / M_\star)$ of $-1.50$, closely matching the values found for Seyferts and IR AGN in our low-redshift sample. Luminous quasars from \citet{Yue2024} and \citet{Ding2023} at $z > 6$ show more variation, with median values of $-0.75$ and $-2.10$, respectively (top panel of Fig. \ref{fig_ratio}). While the Yue et al. sample appears to host relatively overmassive black holes, the Ding et al. quasars are more consistent with the X-ray-selected AGN in our sample. These results suggest that a high black hole-to-stellar mass ratio is a generic feature of low-mass or early-phase systems, persisting across cosmic time. The least-squares fit to the \cite{Sun2025b} sample reinforces this point, with a similar slope and normalization to the low-redshift dwarf Seyferts (bottom panel of Fig. \ref{fig_ratio}).

The observed flatness of the $\log(M_{\mathrm{BH}} / M_\star)$ ratio with redshift for dwarf galaxy AGN contrasts with some theoretical expectations that predict evolving scaling relations due to differential growth of black holes and their host galaxies over cosmic time. Semi-analytic and cosmological hydrodynamic simulations that incorporate AGN feedback and hierarchical assembly often anticipate a gradual decrease in the BH-to-stellar mass ratio at higher redshifts, particularly in low-mass systems where stellar mass builds up faster than black hole mass \citep[e.g.,][]{Volonteri2016, Habouzit2021, AnglesAlcazar2017, Trinca2022, Koudmani2024}. These models suggest that black hole growth in low-mass halos is either delayed by inefficient accretion or suppressed by stellar or AGN feedback. In contrast, our results indicate that within the redshift range probed by our sample ($0.5 < z < 0.9$), dwarf AGN exhibit elevated mass ratios that remain approximately constant with redshift. This apparent stability may challenge models that rely on delayed seeding, low accretion efficiencies, or strong feedback in the early stages of dwarf galaxy evolution.

In the bottom panel of Fig.~\ref{fig_ratio}, we present the $\log(M_{\mathrm{BH}} / M_\star)$ ratio as a function of stellar mass. The dashed lines correspond to least-squares fits applied to the binned data for the three AGN populations in our study, as well as to the individual measurements of the high-redshift Seyferts from \citet{Sun2025b}. Seyferts and IR AGN exhibit a clear decline in $\log(M_{\mathrm{BH}} / M_\star)$ with increasing $M_\star$, while the trend for LINERs is less robust and may be influenced by a single low-mass bin. A similar trend (decline) is observed for the high-redshift Seyferts in dwarf galaxies reported by \citet{Sun2025b}. Interestingly, X-ray-selected AGN from the XXL survey at $z < 2$ also follow this declining trend for $\log(M_\star/M_\odot) < 11$. At higher stellar masses, however, the $\log(M_{\mathrm{BH}} / M_\star)$ ratio flattens and reaches values consistent with those observed in local, inactive galaxies (dotted line), both for the XXL and 4XMM AGN samples. This decline may reflect an early phase of accelerated black hole growth relative to the host galaxy, followed by stellar mass assembly that gradually reduces the M$_\mathrm{BH}$/M$_\star$ ratio toward values typical of local, massive, inactive galaxies. While the data may suggest a decreasing trend in M$_\mathrm{BH}$/M$_\star$ with increasing M$_\star$, particularly when comparing the XMM-XXL and 4XMM samples, we emphasize that this trend remains tentative due to differences in AGN selection and the inherent uncertainties in black hole mass estimates. Nevertheless, if this trend is real, the proposed interpretation aligns with predictions from cosmological simulations, which show that black holes in low-mass halos tend to grow rapidly at early times, while the build-up of stellar mass continues over longer timescales \citep[e.g.,][]{Habouzit2017, AnglesAlcazar2017, Dubois2021}. For LINERs, however, the trend is less clear, with limited dynamic range and larger scatter possibly masking any underlying correlation. At higher stellar masses ($\log(M_\star/M_\odot) \gtrsim 10.8$), the relation flattens and converges toward the canonical local value of $\log(M_{\mathrm{BH}} / M_\star) \sim -3$.

When incorporating $M_{\mathrm{BH}}$ estimates from X-ray AGN and optical quasars, we emphasize that these typically include broad-line (Type 1) sources, for which M$_{\mathrm{BH}}$ is derived using single-epoch virial estimators based on the widths of broad emission lines. These virial methods probe the dynamics of the broad-line region and are fundamentally different from the narrow-line diagnostics used in our Seyfert, LINER, and IR AGN sample. As such, direct comparisons between the two must be interpreted with caution, as each method carries its own systematic uncertainties and probes different physical scales. The scatter and possible offsets in M$_{\mathrm{BH}}$–M$_\star$ space may in part reflect these methodological differences, in addition to any underlying physical diversity in AGN fueling or host galaxy properties.

Taken together, our results suggest that the diversity in AGN populations hosted by dwarf galaxies cannot be explained solely by a single evolutionary sequence. If IR AGN, Seyferts, and LINERs represented consecutive phases of black hole growth (e.g., IR AGN → Seyferts → LINERs), one would expect a smooth increase in stellar mass and perhaps black hole mass along this progression. However, our findings challenge this view. Instead, they point toward a bimodal framework, in which different AGN types may arise from distinct physical conditions and triggering mechanisms. IR-selected AGN may trace early, dust-enshrouded black hole growth episodes—possibly merger-driven and embedded in gas-rich environments. Seyferts could represent ongoing SMBH fueling via secular processes or disk instabilities, consistent with their moderate-to-high accretion rates and elevated star formation. LINERs, in contrast, appear to be hosted by quenched, massive galaxies with older stellar populations and may result from radiatively inefficient accretion, potentially fueled by hot halo gas or residual material from earlier activity. While these trends can be interpreted as snapshots along an evolutionary pathway, they may also reflect fundamental differences in origin, where Seyferts and LINERs arise from distinct fueling modes and galaxy environments \citep[e.g.,][]{Mountrichas2024e}.

\section{Summary}

In this work, we analyzed a sample of 787 Seyfert galaxies, 263 LINERs, and 1\,058 star-forming galaxies selected using the BPT diagnostic diagram, as well as 393 infrared (IR) AGN identified through WISE mid-infrared color selection, which fall outside the BPT classification. All AGN live in dwarf galaxies with $\log(M_\star/M_\odot)<10$, are drawn from the VIPERS catalogue and span the redshift range $0.5 < z < 0.9$ \citep[see][for sample selection details]{Siudek2023a}.

To quantify the impact of AGN activity on star formation, we employed the SFR$_{\mathrm{norm}}$ parameter, defined as the ratio of the SFR in an AGN host to that of a star-forming, non-AGN galaxy of similar stellar mass and redshift. We explored how SFR$_{\mathrm{norm}}$ varies as a function of AGN power (using [OIII] luminosity as a proxy), black hole mass, and large-scale environment (via local overdensity measurements). We also incorporated additional host galaxy properties, including the 4000\AA \,break strength (D${4000}$), to place AGN activity within the broader context of galaxy structure and evolution. Based on this multi-dimensional analysis, we draw the following main conclusions:

\begin{itemize}

    \item All AGN populations exhibit suppressed star formation (SFR$_{\mathrm{norm}} < 1$) at low [OIII] luminosities, relative to matched star-forming galaxies. This suggests AGN activity may suppress star formation in this regime, or that AGN are preferentially found in already quenching galaxies.
    
    \item SFR${_\mathrm{norm}}$ increases with L[OIII] for all AGN types, crossing unity at different luminosity thresholds. This pattern may reflect differences in accretion modes, feedback strength, or selection biases across populations.
    
    \item SFR$_{\mathrm{norm}}$ shows distinct behaviors across AGN populations as a function of black hole mass: LINERs display an approximately flat trend with values slightly below but broadly consistent with unity; Seyferts exhibit a mild increase in SFR$_{\mathrm{norm}}$ with M$_{\mathrm{BH}}$; and IR-selected AGN show a more pronounced upward trend. For both IR and X-ray AGN, SFR$_{\mathrm{norm}}$ exceeds unity at higher black hole masses, suggesting enhanced star formation in these systems.

    \item SFR$_{\mathrm{norm}}$ remains relatively flat across all environmental densities for Seyferts, with values slightly below unity. LINERs and IR AGN show a possible rise in SFR$_{\mathrm{norm}}$ at the highest overdensities, but this is driven by a single bin and should be interpreted with caution. Overall, there is no strong evidence for a systematic dependence of SFR$_{\mathrm{norm}}$ on environment in any AGN population.

    \item LINERs and IR AGN exhibit slightly older stellar populations (median D${4000}$ $\sim$ 1.18–1.19) compared to Seyferts (D${4000} \sim$ 1.14).

\end{itemize}

  We further examine the relationship between black hole mass and stellar mass, as well as the evolution of their ratio as a function of redshift and stellar mass across the three AGN populations. To place our findings in a broader context, we compare them with results from previous X-ray selected AGN studies and incorporate high-redshift ($z \sim 4$–7) AGN samples. This allows us to extend the analysis toward more massive galaxies ($10 < \log(M_\star/M_\odot) < 12$) and earlier cosmic epochs, providing a unified view of black hole–galaxy co-evolution across a wide range of mass and redshift. The main conclusions from this part of our analysis are as follows:

\begin{itemize}

    \item At fixed stellar mass, Seyfert galaxies tend to host more massive black holes than LINERs, resulting in higher median $\log(M_{\mathrm{BH}} / M_\star)$ ratios, with IR AGN falling in between. Compared to our AGN populations, X-ray-selected AGN from the 4XMM and XMM-XXL surveys show systematically lower $\log(M_{\mathrm{BH}} / M_\star)$ ratios, suggesting differences in selection or evolutionary stage, and highlighting the importance of AGN classification in interpreting black hole–host scaling relations.

    \item LINERs exhibit the steepest slope in the $M_{\mathrm{BH}}$–$M_\star$ relation among all AGN populations, suggesting a stronger coupling between black hole and stellar mass in these sources, reminiscent of bulge-dominated, quenched galaxies.

    \item Seyferts and IR AGN in dwarf galaxies show a declining trend in $\log(M_{\mathrm{BH}} / M_\star)$ with increasing stellar mass, consistent with a scenario in which black hole growth initially outpaces stellar mass assembly in low-mass systems. X-ray AGN at similar redshifts exhibit a comparable decline up to $\log(M_\star/M_\odot) \sim 11$, above which the ratio flattens toward values typical of local, inactive galaxies. In contrast, LINERs do not show a clear trend, possibly due to limited dynamic range or larger scatter in the data.

    \item High-redshift ($z > 4$) AGN, including Seyferts from \citet{Sun2025} and quasars from \citet{Ding2023} and \citet{Yue2024}, show $\log(M_{\mathrm{BH}} / M_\star)$ values comparable to those of our dwarf AGN populations, suggesting that elevated black hole mass fractions are not unique to the local universe but extend to the early phases of galaxy evolution.
    
\end{itemize}

In summary, our joint analysis of star formation activity, black hole mass, and host galaxy properties reveals a diverse set of AGN evolutionary pathways within dwarf galaxies. This diversity likely reflects a combination of physical processes. For instance, LINERs exhibit older stellar populations and lower SFR$_\mathrm{norm}$ values, consistent with evolved systems with limited cold gas supply. In contrast, Seyferts show higher black hole masses at fixed stellar mass and elevated SFR$_\mathrm{norm}$, suggesting more active black hole growth in gas-rich environments. IR-selected AGN, with their flatter M$_\mathrm{BH}$–M$_\star$ relation and intermediate D4000 values, may represent obscured systems undergoing a transitional phase of growth. These differences point to a complex interplay between gas availability, obscuration, feedback strength, and environmental factors. The use of complementary AGN selection techniques, each probing different accretion phases, underscores the need for a multi-faceted approach to fully capture the range of AGN activity. Future spectroscopic and multiwavelength surveys, particularly those enabling morphological and spatially resolved analysis, such as DESI, will provide critical constraints on these evolutionary scenarios across broader parameter space.

\begin{acknowledgements}
GM acknowledges funding from grant PID2021-122955OB-C41 funded by MCIN/AEI/10.13039/501100011033 and by “ERDF/EU”. M.S. acknowledges support by the State Research Agency of the Spanish Ministry of Science and Innovation under the grants 'Galaxy Evolution with Artificial Intelligence' (PGC2018-100852-A-I00) and 'BASALT' (PID2021-126838NB-I00) and the Polish National Agency for Academic Exchange (Bekker grant BPN/BEK/2021/1/00298/DEC/1). This work was partially supported by the European Union's Horizon 2020 Research and Innovation program under the Maria Sklodowska-Curie grant agreement (No. 754510).

\end{acknowledgements}

\bibliography{mybib}
\bibliographystyle{aa}

\end{document}